\documentclass[preprint,authoryear,5p,twocolumn,12pt]{elsarticle}

\usepackage{amsmath}
\usepackage{amssymb}
\newcommand{\degree}{\ensuremath{^\circ}}
\newcommand\T{\rule{0pt}{3ex}}       
\newcommand\B{\rule[-1.5ex]{0pt}{0pt}} 

\journal{Planetary and Space Science}

\begin{document}

\begin{frontmatter}



\title{The primordial collisional history of Vesta: crater saturation, surface evolution and survival of the basaltic crust}


\author{D. Turrini}
\fntext[1]{Email: diego.turrini@iaps.inaf.it\\ Phone: +390649934414 \\ Fax: +390649934383}
\address{Istituto di Astrofisica e Planetologia Spaziali INAF-IAPS, Via del Fosso del Cavaliere 100, 00133, Rome, Italy}

\begin{abstract}
The Dawn mission recently visited the asteroid $4$ Vesta and the observations performed by the spacecraft revealed more pieces of the intriguing mosaic of its history. Among the first results obtained by the Dawn mission was the confirmation of the link between the howardite - eucrite - diogenite (HED) group of meteorites and Vesta. This link implies that Vesta was one of the first objects to form in the Solar System and that the differentiation of the asteroid likely completed before the formation of Jupiter. As a consequence, the bombardment triggered by the formation and migration of the giant planet, the Jovian Early Bombardment (JEB), contributed to the collisional evolution of the asteroid at a time where most of its interior was still molten. This work explores the implications of the JEB for the evolution of the primordial Vesta, in particular in terms of crater saturation, crustal excavation and surface erosion. Both scenarios assuming the planetesimals having formed in a quiescent or a turbulent nebula were explored and both primordial and collisionally evolved size-frequency distributions were considered. 
The results obtained indicate that, if the basaltic surface of Vesta were already formed, the JEB would saturate it with craters and could erode it to depths that vary from hundreds of meters to tens of kilometres. In the latter cases, the surface erosion caused by the JEB would be comparable with the thickness of the eucritic and diogenitic layers of Vesta. In the cases where the global surface erosion is limited, however, large impactors, if too abundant, can excavate the whole crust and extract significant quantities of material from the vestan mantle, incompatible with the present understanding of HED meteorites. This appears to be the case if the impacting planetesimals formed in a turbulent nebula and Jupiter migrated by $0.5$ AU or more. Globally, the results obtained suggest that the scenarios where the planetesimal formed in a quiescent nebula and Jupiter underwent a modest migration (i.e. up to $0.5$ AU) are the most consistent with our understanding of Vesta, even if the cases of planetesimals formed in a turbulent nebula with Jupiter undergoing limited (i.e. about $0.25$ AU) or no migration cannot be ruled out. Recent results on the differentiation of the asteroid, however, raised the possibility that Vesta originally possessed a now-lost undifferentiated crust. In this case, the favoured scenarios would be those where the planetesimals formed in a quiescent nebula and Jupiter underwent a more significant migration (i.e. between $0.5$ AU and $1$ AU).

\end{abstract}


\end{frontmatter}


\section{Introduction}
\label{intro}

More than 40 years ago, the asteroid $4$ Vesta was suggested, based on spectroscopic measurements, to be the parent body of the Howardite-Eucrite-Diogenite (HED in the following) class of meteorites \citep{mccord1970}. The Vesta-HED link was later supported by the observations of the Hubble Space Telescope \citep{gaffey1997,binzel1997,li2010} and was recently confirmed by the observations of the Dawn spacecraft \citep{desanctis2012a,prettyman2012}, which spent one year gathering data on the composition, morphology and gravity of the asteroid \citep{russell2012,russell2013}. The Vesta-HED link allows the use of the radiometric ages of HED meteorites to investigate the ancient past of the asteroid. As a consequence, based on the crystallization ages of the oldest eucrites \citep{bizzarro2005} and diogenites \citep{schiller2011} we know that Vesta formed and differentiated in the first $3$ Ma of the Solar System's lifetime.

At the time Vesta was forming and differentiating, the Solar System was in the phase of its evolution known as the Solar Nebula \citep{coradini2011}, i.e. it was a circumsolar disk of gas and dust where the first generations of planetary bodies were forming. The beginning of the Solar Nebula is generally assumed to coincide with the condensation of the oldest solids, the Ca-Al-rich inclusions (CAIs), about $4.568$ Ga ago \citep{bouvier2010}. The duration of the Solar Nebula phase is indirectly constrained by observations of circumstellar disks, which indicate that the median lifetime is about $3$ Ma with the range of observed values spanning between $1-10$ Ma \citep{meyer2009}. We know that the giant planets formed in the Solar Nebula, as the nebular gas supplied the material for the massive envelopes of Jupiter and Saturn and the limited ones of Uranus and Neptune. Theoretical \citep{bottke2005a,bottke2005b} and observational \citep{scott2006} arguments suggests that Jupiter formed $3-5$ Ma after the condensation of CAIs. As a consequence, the differentiation and the formation of the basaltic crust of Vesta predate the formation of Jupiter, plausibly the first giant planet to have formed \citep{coradini2011}.

While we know that Vesta is differentiated and possesses an iron core \citep{russell2012,russell2013}, its internal structure is still a matter of debate. Depending on the initial composition of the asteroid, \citet{ruzicka1997} estimated that the global thickness of the crust from which eucrites and diogenites originated should have ranged between $40$ km and $80$ km. More specifically, \citet{ruzicka1997} report that the eucritic and diogenitic layers would have thickness of $26$ km and $13$ km, respectively, if Vesta had an initial composition similar to CI meteorites (i.e. olivine-rich). If the material initially composing Vesta was instead more similar to EH meteorites (i.e. olivine-poor), the thickness of the eucritic and diogenitic layers would respectively be $41$ km and $42$ km. However, \citet{mcsween2013} argues that the latter case likely assumes an unrealistic initial composition of Vesta and favour the former scenario. 

It is important to note that the ages of the oldest HED meteorites \citep{bizzarro2005,schiller2011} indicate that eucritic and diogenitic material was already crystallizing at $3$ Ma, but this does not necessarily imply that the whole basaltic crust of Vesta was completely solidified. Given the large range of temperatures over which silicates are partially molten (more than $400$ K), it is more than likely that over a temporal window of at least a few Ma after the crystallization of the oldest samples the crust of Vesta still contained molten material. The results of thermal models (see e.g. \citealt{formisano2013} and references therein for previous works on the subject by other authors) and geophysical models \citep{tkalcec2013} point in this direction, indicating that between $3$ Ma and $5$ Ma from CAIs the thickness of the completely solidified crust could have gone from a minimum of $7$ km (\citealt{formisano2013,tkalcec2013}, Supplementary Information) to about $30$ km \citep{formisano2013}.

As first pointed out by \citet{davis1985}, the survival of the basaltic crust of Vesta to the collisional evolution the asteroid underwent across the lifetime of the Solar System represents one of the stronger constraints to understand the past history of the asteroid belt. The present surface composition of Vesta observed by Dawn \citep{desanctis2012a,prettyman2012} is globally consistent with howardites (breccias formed by a mixture of eucritic and diogenitic material), with regions more similar to eucrites and with an exposure of diogenitic material in the south polar basin RheaSilvia \citep{desanctis2012a,prettyman2012,mcsween2013}. This implies that the surface erosion of Vesta should not exceed the values estimated by \citet{ruzicka1997}. The southern hemisphere of Vesta, however, was extensively excavated by the impacts that generated the RheaSilvia basin and the underlying Veneneia basin \citep{schenk2012,jutzi2013,mcsween2013,ivanov2013}. The global erosion experienced by the surface of Vesta from the formation of its basaltic crust to the formation of these two basins is therefore an important piece in the mosaic of the history of this body and the asteroid belt in general.

After the formation of Jupiter and Saturn, the asteroid belt underwent a rapid process of depletion due to the interplay between the orbital resonances created by the two giant planets and the gravitational perturbations of the planetary embryos in the inner Solar System (see \citealt{coradini2011,obrien2011} and references therein). Following this phase of depletion, the population of the asteroid belt decreased by about two orders of magnitude (see \citealt{coradini2011,obrien2011} and references therein). \citet{bottke2005a,bottke2005b} investigated the collisional evolution of the asteroid belt from the beginning of this depletion process to present time and found that it is globally consistent with the survival of the basaltic crust of Vesta. In these studies, the authors did not account for the effects of cratering erosion but focus on those of catastrophic disruption. According to \citet{davis1979}, if the primordial population of the asteroid belt was very steep, cratering erosion would contribute to the global collisional evolution of the asteroid belt and dominate that of small bodies with negligible gravity. Even in this case, however, the asteroid belt would naturally evolve to a more relaxed state where cratering erosion is not important in less than $10^{8}$ years \citep{davis1979}, i.e. on a timescale much shorter that the one considered by \citet{bottke2005a,bottke2005b}.

A temporal interval not covered by the studies of \citet{bottke2005a,bottke2005b} is the one going from the beginning of the formation of Jupiter to the end of that of Saturn.  The formation of Jupiter has been shown by different authors \citep{safronov1972,weidenschilling1975,weidenschilling2001,turrini2011,turrini2012} to trigger a sudden spike in the flux of impactors in the early history of the Solar System. This event, named the Jovian Early Bombardment (\citealt{turrini2011,turrini2012}, JEB in the following), is caused by the scattering of ice-rich planetesimals from the outer Solar System due to the gravitational perturbation of the giant planet \citep{safronov1972,weidenschilling1975,weidenschilling2001,turrini2011,turrini2012} and by the appearance of the Jovian mean motion resonances in the asteroid belt, in particular the $3$:$1$ and $2$:$1$ resonances \citep{weidenschilling2001,turrini2011,turrini2012}. The duration of the JEB is limited to about $1$ Ma \citep{weidenschilling1975,turrini2011,turrini2012}, with the bulk of the impacts taking place in the first $3-5\times10^{5}$ years \citep{turrini2011}. The flux of impactors due to the Jovian resonances is the dominant one in the inner Solar System \citep{turrini2011} and is the one shaping the early collisional evolution of the asteroid belt \citep{turrini2011,turrini2012}.

\citet{turrini2011} estimated the fluxes of impactors coming from the outer Solar System and from the Jovian resonances during the JEB, the crater populations they produce and the probability of Vesta being destroyed during the bombardment using different size-frequency distributions (SFDs in the following) of the impactors. Their results showed that the probability of Vesta undergoing a catastrophic impact are negligible, but suggested that cratering erosion could play an unexpectedly significant role due to the higher, pre-depletion population of planetesimals inhabiting the asteroid belt at the time. \citet{turrini2012} further investigated the subject of asteroidal erosion during the JEB and, using a more detailed physical description of the mass loss processes, showed that cratering erosion indeed played a much more relevant role than catastrophic disruption in determining the fate of primordial asteroids. \citet{turrini2012} showed that cratering erosion is a function of the extent of Jupiter's migration and of the position of the target body in the asteroid belt. Depending on the considered scenario and SFD of the impactors, planetesimals the size of Vesta could lose from a few times $1\%$ to a few times $10\%$ of their original mass \citep{turrini2012}. 

This work will discuss the primordial surface evolution of Vesta due to the JEB and extend the analysis of the results described in \citet{turrini2011}, by reprocessing them with an updated version of the collisional model first applied in \citet{turrini2012} and by putting them in the context of the most updated understanding of the geophysical state of Vesta at that time. Together with the SFDs already considered in \citet{turrini2011,turrini2012}, this work will include in the analysis the SFD for primordial asteroids formed from small ($50-500$ m in diameter) planetesimals proposed by \citet{weidenschilling2011}. Before proceeding, it is important to note that the original simulations of \citet{turrini2011} focused on the role of Jupiter and neglected such effects like gas drag and the gravitational perturbations of the planetary embryos already existing at the time. As a consequence, the goal of this work is to provide a first exploration of the implications of the JEB for the survival of the basaltic crust of Vesta and to pave the road for future, more detailed studies.

\section{Jovian Early Bombardment and collisional model}
\label{model}

The input data used to investigate the surface evolution of Vesta across the JEB in this work are mainly the fluxes of impactors described in \citet{turrini2011}, hereafter Paper I. As the dynamical, physical and numerical details of the model used to study the JEB have been extensively described in Paper I and in \citet{turrini2012}, hereafter Paper II, only the main aspects of the simulations will be highlighted here. The interested readers are referred to Papers I and II for more details. 

\subsection{Solar Nebula}

The template of the Solar Nebula at the beginning of the simulations was composed of the Sun, Vesta, the forming Jupiter and a disk of planetesimals modeled as massless particles. The evolution of this template of the Solar System was followed for $2\times10^{6}$ years, centered on the time Jupiter reached its final mass.

At the beginning of the simulations, Jupiter was a planetary embryo with mass $M_{0}=0.1\,M_\oplus$ that grew to the critical mass $M_{c}=15\,M_\oplus$ in $\tau_{c}=10^{6}$ years as:
\begin{equation*}
 M=M_{0}+\left( \frac{e}{e-1}\right)\left(M_{c}-M_{0}\right)\times\left( 1-e^{-t/\tau_{c}} \right)
\end{equation*}
As shown in Papers I and Paper II, the effects of Jupiter on the collisional evolution of Vesta during this phase are negligible. The Jovian mass is too low to perturb the asteroid belt and asteroidal impacts on Vesta result in net accretion. A few bodies in the outer Solar System can have close encounters with Jupiter, be injected on orbits crossing the asteroid belt and impact on Vesta, but they are not considered in this analysis.

When the critical mass value $M_{c}$ is reached, the nebular gas surrounding Jupiter is assumed to be rapidly accreted by the planet, whose mass then grows as:
\begin{equation*}
 M=M_{c}+\left( M_{J} - M_{c}\right)\times\left( 1-e^{-(t-\tau_{c})/\tau_{g}}\right)
\end{equation*}
where $M_{J}=1.8986\times10^{30}\,g=317.83\,M_{\oplus}$ is the final mass of the giant planet. The e-folding time $\tau_{g}=5\times10^3$ years is derived from the hydrodynamical simulations described in \citet{lissauer2009} and \citet{coradini2010}. While accreting the nebular gas, Jupiter migrated inward due to disk-planet interactions as:
\begin{equation*}
 r_{p}=r_{0}+\left( r_{J} - r_{0}\right)\times\left( 1-e^{-(t-\tau_{c})/\tau_{r}}\right)
\end{equation*}
where $r_{0}$ is the orbital radius of Jupiter at the beginning of the simulation, $r_{J}$ is the present orbital radius and $\tau_{r}=\tau_{g}=5\times10^{3}$ years. Paper I considered four different migration scenarios: no displacement and $0.25$ AU, $0.5$ AU and $1$ AU displacements. The initial position of Jupiter is chosen in the different scenarios so that its final position is always the present one, consistently with the original Nice model \citep{tsiganis2005} but not with its most recent developments \citep{levison2011} where Jupiter should be at about $5.4$ AU. With the chosen values of $\tau_{r}$, the migration scenarios are equivalent to assume values of $a/\dot{a}$ respectively equal to $3.2\times10^{5}$ years, $1.6\times10^{5}$ years and $8\times10^{4}$ years, consistently with the results of theoretical studies \citep{papaloizou2007}. As a test, Paper I modified the values of $\tau_{g}$ and $\tau_{r}$ to $2.5\times10^{4}$ years in the scenario were Jupiter migrates by $1$ AU, but neither the fluxes of impactors nor the impact velocities on Vesta changed in any significant way (the same is not true for Ceres due to its position between the 3:1 and the 2:1 resonances).

Vesta was initially placed on a circular, planar orbit with semimajor axis $a_{v}=2.362$ AU and dynamically evolved under the effect of the Jovian perturbations. The asteroid was characterized by a mean radius $r_{v}=258$ km \citep{thomas1997} and a mass $m_{v}=2.70\times10^{23}$ g \citep{michalak2000}. Note that these values of the mean radius and the mass of Vesta are slightly different from the ones estimated by the Dawn mission ($262.7$ km and $2.59\times10^{23}$ g respectively, \citealt{russell2012}). However, these values were only used to evaluate the collisional cross-section of Vesta and the differences between the two sets of values are of the order of only $2-4\%$, so they do not significantly affect the fluxes estimated in Paper I. The average density of Vesta is $\overline\rho_{v}=3.456$ g\,cm$^{-3}$ \citep{russell2012} and the density of its crust, which we used in the collisional model, is $\rho_{v}=3.090$ g\,cm$^{-3}$ \citep{russell2012,russell2013}.

The disk of planetesimals extended between $2$ AU and $10$ AU, and the planetesimals had initial values of the orbital eccentricity and inclination ranging between $0 \leq e_{i} \leq 3\times10^{-2}$ and $0$ rad $\leq i_{i} \leq 3\times10^{-2}$ rad, respectively. Planetesimals formed in the inner Solar System (ISS in the following) were considered rocky bodies with mean density $\rho_{iss}=3.0$ g cm$^{-3}$, while planetesimals formed in the outer Solar System (OSS in the following) were considered volatile-rich bodies with mean density $\rho_{oss}=1.0$ g cm$^{-3}$. Differently from Paper I, the change between the inner and the outer Solar System in this work was always assumed to occur at $r_{SL}=4.0$ AU, assumed to be the location of the Snow Line. The planetesimals are removed from the simulations if their semimajor axes become smaller than $1$ AU or larger than $30$ AU and if they impact the Sun, Jupiter or Vesta. As mentioned previously, in the simulations of Paper I the effects of gas drag on the orbital evolution of the planetesimals were not included. Note, however, that in the study of \citet{weidenschilling2001}, which includes the effects of gas drag, the OSS resonant population of impactors influenced by the $3$:$2$ and $2$:$1$ resonances still exists. 

The density values of the planetesimals, together with those of the mass and the normalization factors described in Sect.\ref{collisions}, were used to estimate the collisional evolution of Vesta through Monte Carlo simulations as detailed in the following.\\

\subsection{Collisional model}
\label{collisions}

To estimate the fluxes of impactors on Vesta, Paper I opted for a statistical approach based on solving the ray--torus intersection problem between the orbital torus of Vesta and the linearized path of a massless particle across a time step. The method is similar to the analytical method developed by \citet{opik1976}, but does not require averaging over orbital angles other than the mean anomaly. Interested readers are referred to Paper I and II for details on the algorithm.

The fluxes estimated in Paper I were used as the basis for an improved assessment of the collisional evolution of Vesta during the JEB. A set of $10^{4}$ Monte Carlo simulations was run for each of the planetesimal SFDs described in Sect. \ref{distributions}. In each run a new mass value was extracted for each impact event recorded in the simulations of Paper I and, using the corresponding impact velocity, the diameter of the produced crater, the energy of the impact event, and the eroded mass were computed. Averaging over each set of $10^{4}$ Monte Carlo simulations, for each SFD of the primordial planetesimals the total eroded mass, the cumulative probability of Vesta undergoing catastrophic disruption, and the fraction of the Vestan surface affected by the impacts were then computed.

The diameter of the craters produced by the flux of impactors was estimated using the following scaling law for rocky targets by \citet{holsapple2007}:
\begin{align}\label{crater_law}
\frac{R_{c}}{r_{i}}=0.93\left(\frac{g\,r_{i}}{v_{i}^{2}}\right)^{-0.22}	\left(\frac{\rho_{i}}{\rho_{v}}\right)^{0.31}\nonumber\\
+0.93\left(\frac{Y_{v}}{\rho_{v} v_{i}^2}\right)^{-0.275}\left(\frac{\rho_{i}}{\rho_{v}}\right)^{0.4}
\end{align}
where $R_{c}$ is the final radius of the crater, $r_{i}$ is the radius of the impactor, $g=0.25$ m\,s$^{-1}$ is the surface gravity of Vesta, $v_{i}$ is the impact velocity, $Y_{v}=7.6$ MPa is the strength of the material composing the surface of Vesta (assumed to behave as soft rock, \citealt{holsapple1993}), $\rho_{i}$ and $\rho_{v}$ are the densities respectively of the impactor and of the basaltic surface of Vesta.

The specific energy $Q_{D}$ of the impact events is expressed in units of the specific dispersion energy $Q^{*}_{D}$ of the target body \citep{benz1999}. This study evaluates the catastrophic disruption threshold $Q^{*}_{D}$ of Vesta using Eq. $6$ from \citet{benz1999} and the coefficients for basaltic targets computed by these authors (see Table $3$, ibid). As in Paper II, the coefficients of the case $v_{i}=5\,km\,s^{-1}$ were used for all impact events with a velocity greater or equal than $5\,km\,s^{-1}$, and those of the $v_{i}=3\,km\,s^{-1}$ were used for all the other impact events.

To evaluate the surface erosion of Vesta, this study proceeded in a way analogous to that of Paper II. For low-energy impacts ($Q_{D}/Q^{*}_{D} < 0.1$), this study used the scaling law for rocky targets by \citet{holsapple2007} in the form developed by \citet{svetsov2011} averaging over all impact angles:
\begin{equation}\label{erosion_svetsov}
\frac{m_{e}}{m_{i}}=0.03\left(\frac{v_{i}}{v_{V}}\right)^{1.65}\left(\frac{\rho_{i}}{\rho_{V}}\right)^{0.2}
\end{equation}
where $m_{e}$ is the escaped mass, $m_{i}$ is the mass of the impactor, $v_{i}$ is the impact velocity, $v_{V}$ is the escape velocity from the surface of Vesta and $\rho_{i}$ and $\rho_{V}$ are the densities respectively of the impactor and of Vesta.
For high-energy impacts ($0.1 \leq Q_{D}/Q^{*}_{D} < 1$), this study used instead Eq. $8$ from \citet{benz1999} expressed in terms of the eroded mass: 
\begin{equation}\label{erosion_benz}
\frac{m_{e}}{m_{t}}=0.5+s\left(\frac{Q_{D}}{Q^{*}_{D}}-1.0\right)
\end{equation}
where $s=0.5$ for $v_{i}<5\,km\,s^{-1}$ and $s=0.35$ for $v_{i} \geq 5\,km\,s^{-1}$. The effects of catastrophic impacts ($Q_{D}/Q^{*}_{D} \geq 1$) were not accounted for in the estimates of the eroded mass. The cumulative number of catastrophic impacts was used only to assess the probability of Vesta surviving the JEB. 

\subsection{Size-frequency distributions of planetesimals}
\label{distributions}

This study considered a total of $5$ SFDs of the primordial planetesimals populating the Solar Nebula, expanding the set of SFDs investigated in Paper II with the one recently published by \citet{weidenschilling2011}.
\subsubsection*{Planetesimals formed in a quiescent disk}
The first SFD considered was that of a disk of planetesimals formed by gravitational instability of the dust in the mid-plane of a non turbulent protoplanetary nebula \citep{safronov1972,goldreich1973,weidenschilling1980}. The protoplanetary nebula was assumed to have a mass of $M_{neb}=0.02$ M$_\odot$ distributed between $1-40$ AU with dust-to-gas ratio $\xi=0.01$ and density profile $\sigma=\sigma_{0}\left(\frac{r}{1\,AU}\right)^{-n_{s}}$, where $\sigma_{0}=2700$ g cm$^{-2}$ is the surface density at $1$ AU and $n_{s}=1.5$. The initial mass of solids contained in the region comprised between $2$ and $3$ AU (i.e. the reference region considered also by \citealt{morbidelli2009} and \citealt{weidenschilling2011}) is about $4$ M$_{\oplus}$. For such a nebula it can be showed \citep{coradini1981} that the average mass of the planetesimals would follow the semi-empirical relationship
\begin{equation}\label{masslaw}
 \overline{m}_{p}=m_{0}\left( \frac{r}{1\,AU} \right)^{\beta}
\end{equation}
where $\overline{m}_{p}$ and $m_{0}$ are expressed in $g$, $r$ is expressed in $AU$ and $\beta=1.68$. The value $m_{0}$ is the average mass of a planetesimal at $1$ AU, i.e. $2\times10^{17}$ g \citep{coradini1981}. Paper I showed that, assuming that the mass dispersion of the planetesimals about the average values of Eq. \ref{masslaw} is governed by a Maxwell-Boltzmann distribution, a mass value can be associated to each test particle by means of a Monte Carlo method where the uniform random variable $Y$ varying in the range $[0,1]$ is 
\begin{equation}
 Y=\frac{2\gamma\left(3/2,y^{*}\right)}{\sqrt{\pi}}=P\left(3/2,y^{*}\right)
\end{equation}
where $P\left(3/2,y^{*}\right)$ is the lower incomplete Gamma ratio.
The inverse of the lower incomplete Gamma ratio can be computed numerically and, by substituting $y^{*}$ back with $m^{*}/\overline{m}_{p}(r)$ one obtains
\begin{equation}\label{massval}
 m(r)=\overline{m}_{p}inv\left(P\left(3/2,Y\right)\right)
\end{equation}
Since the use of massless particles assured the linearity of the processes investigated over the number of considered bodies, the number of impacts expected in such a disk of planetesimals was extrapolated by multiplying the number of impacts recorded in the simulations of Paper I by a factor $\gamma$ where
\begin{equation}\label{ratio}
 \gamma=N_{tot}/n_{mp}
\end{equation}
where $n_{mp}=8\times10^{4}$ and $N_{tot}$ is given by
\begin{align}\label{ntot}
 N_{tot}=\int^{r_{max}}_{r_{min}} 2 \pi r n^{*}(r)dr= \nonumber \\
 =\pi^{3/2}\frac{\xi\sigma_{0}}{m_{0}}\left(1\,AU\right)^{2}\left(\frac{1}{2-n_{s}-\beta}\right)\times \nonumber \\
 \left(\left(\frac{r_{max}}{1\,AU}\right)^{2-n_{s}-\beta}-\left(\frac{r_{min}}{1\,AU}\right)^{2-n_{s}-\beta}\right)
\end{align}
where $r_{min}=2$ AU, $r_{max}=10$ AU and the symbol $1\,AU$ indicate the value of the astronomical unit expressed in cm.
\subsubsection*{Planetesimals formed in a turbulent disk}
The second SFD considered was that of planetesimals formed by concentration of dust particles in low vorticity regions in a turbulent protoplanetary nebula \citep{cuzzi2008,cuzzi2010}. Following \citet{chambers2010}, the protoplanetary nebula was characterized by a surface density $\sigma'_{0}=3500$ g cm$^{-2}$ at $1$ AU, a nebula density profile with exponent $n'_s=-1$ and a dust-to-gas ratio $\xi'=0.01$ beyond the Snow Line and $\xi'=0.005$ inside the Snow Line (see Fig. $14$, gray dot-dashed line, ibid). Differently from \citet{chambers2010}, also for this SFD the Snow Line was assumed to be at $4$ AU. As in the case of the SFD by \citet{coradini1981}, the initial mass of solids contained in the region comprised between $2$ and $3$ AU is about $4$ M$_{\oplus}$. The results of \cite{chambers2010} supply the average diameter of planetesimals as a function of heliocentric distance (see Fig. $14$, gray dot-dashed line, ibid), from which Paper I derived the following semi-empirical relationship analogous to Eq. \ref{masslaw}:
\begin{equation}\label{chambers_mass}
 \overline{m'}_{p}=\frac{\pi}{6}\rho D_{0}^{3}\left( \frac{r}{1\,AU} \right)^{3\beta'}
\end{equation}
where $\beta'=0.4935$ and $D_{0}=70$ km is the average diameter of the planetesimals at $1$ AU. By substituting the primed quantities to the original ones in Eqs. \ref{massval} and \ref{ntot}, the mass and the normalization factor for each massless particle can be obtained through the same approach described previously.\\
\subsubsection*{The ``Asteroids were born big'' scenario}
The third and the fourth SFDs considered were derived by the results of \citet{morbidelli2009}. \citet{morbidelli2009} did not explore a specific model of planetesimal formation in quiescent or turbulent disks but instead tried to constrain the initial size-frequency distribution of planetesimals in the orbital region of the asteroid belt, assuming a initial mass of $1.6$ M$_{\oplus}$ in the region comprised between $2$ and $3$ AU. Their results suggest that the best match with the present-day SFD of the asteroid belt is obtained for planetesimal sizes initially spanning $100-1000$ km (see Fig. $8$, ibid), a range consistent with their formation in a turbulent nebula. \citet{morbidelli2009} supplies two SFDs associated to this case: a first one describing the primordial SFD of the planetesimals, which spans $100-1000$ km (see Fig. $8a$, black dots, ibid), and a second, collisionally evolved one where accretion and break-up of the primordial planetesimals extended the size distribution between $5-5000$ km (see Fig. $8a$, black solid line, ibid). For each ISS impact event in the simulations of Paper I the mass of the impacting planetesimals was estimated through a simple Monte Carlo extraction based on the cumulative probability distributions of the two SFDs supplied by \cite{morbidelli2009}. The normalization factor was estimated through Eq. \ref{ratio} using the total number of planetesimals in the asteroid belt supplied by the cumulative SFDs from Fig. $8a$ in \cite{morbidelli2009}. The SFDs from \cite{morbidelli2009} are valid only for ISS impactors, since these authors focused on the primordial SFD of planetesimals in the orbital region of the asteroid belt.
\subsubsection*{The ``Asteroids were born small'' scenario}
The final SFD considered was derived from the results of \citet{weidenschilling2011}, who studied the accretion of primordial planetesimals in an annular region comprised between $1.5$ AU and $4$ AU and containing $4.9$ M$_{\oplus}$ ($2$ M$_{\oplus}$ in the region between $2$ AU and $3$ AU). Using an approach analogous to the one also used by \citet{morbidelli2009} but differing in the algorithms governing the computation of the collisional probabilities, \citet{weidenschilling2011} showed that a primordial SFD of the asteroid belt capable of reproducing the features of the present day SFD can be obtained also from disks initially populated by planetesimals as small as $50-200$ m. Planetesimals of $500$ m in diameter succeed only partially in producing a satisfactory SFD. This study focused on the SFD of the asteroid belt that \citet{weidenschilling2011} referred to as the ``standard case'', i.e. the one produced from a disk initially populated by planetesimals with a diameter of $100$ m (see Fig. $8$, ibid). As in the case of the SFDs by \citet{morbidelli2009}, the mass of the impacting planetesimals was estimated through a simple Monte Carlo extraction and the normalization factor $\gamma$ was computed using the total number of planetesimals in the asteroid belt as obtained from the considered cumulative probability distribution. Similarly to the two previous SFDs, also the SFD from \citet{weidenschilling2011} is valid only for impactors originating from the asteroid belt. In the following, when using this SFD only planetesimals whose size is greater than or equal to $1$ km will be considered. Note that in the size range considered, the SFD of the primordial asteroids would not change significantly should the initial mass of the region between $2$ AU and $3$ AU be raised to $4$ M$_{\oplus}$ (see Fig. 19 in \citealt{weidenschilling2011}).

\section{Results}

As anticipated in the previous sections, the analysis of the primordial collisional evolution of Vesta focused on the following aspects: the crater saturation of the surface, the survival of Vesta to the JEB, the erosion of Vesta and the survival of its basaltic crust. For the latter, the discussion of the results will consider both the case where it is completely solid, as argued by \citealt{schiller2011}, and the one where it is still characterized by the presence of significant quantities of molten material, as in the theoretical models of \citealt{formisano2013} and \citet{tkalcec2013}. As the SFDs from \citet{morbidelli2009} and \citet{weidenschilling2011} are not valid for OSS impactors, this study focused only on the effects of ISS impactors but took advantage of the results of Paper II to include the effects of the OSS impactors when necessary. 

\subsection{Impact velocities and impactors on Vesta}

\begin{figure*}[t]
\centering
\includegraphics[width=\textwidth]{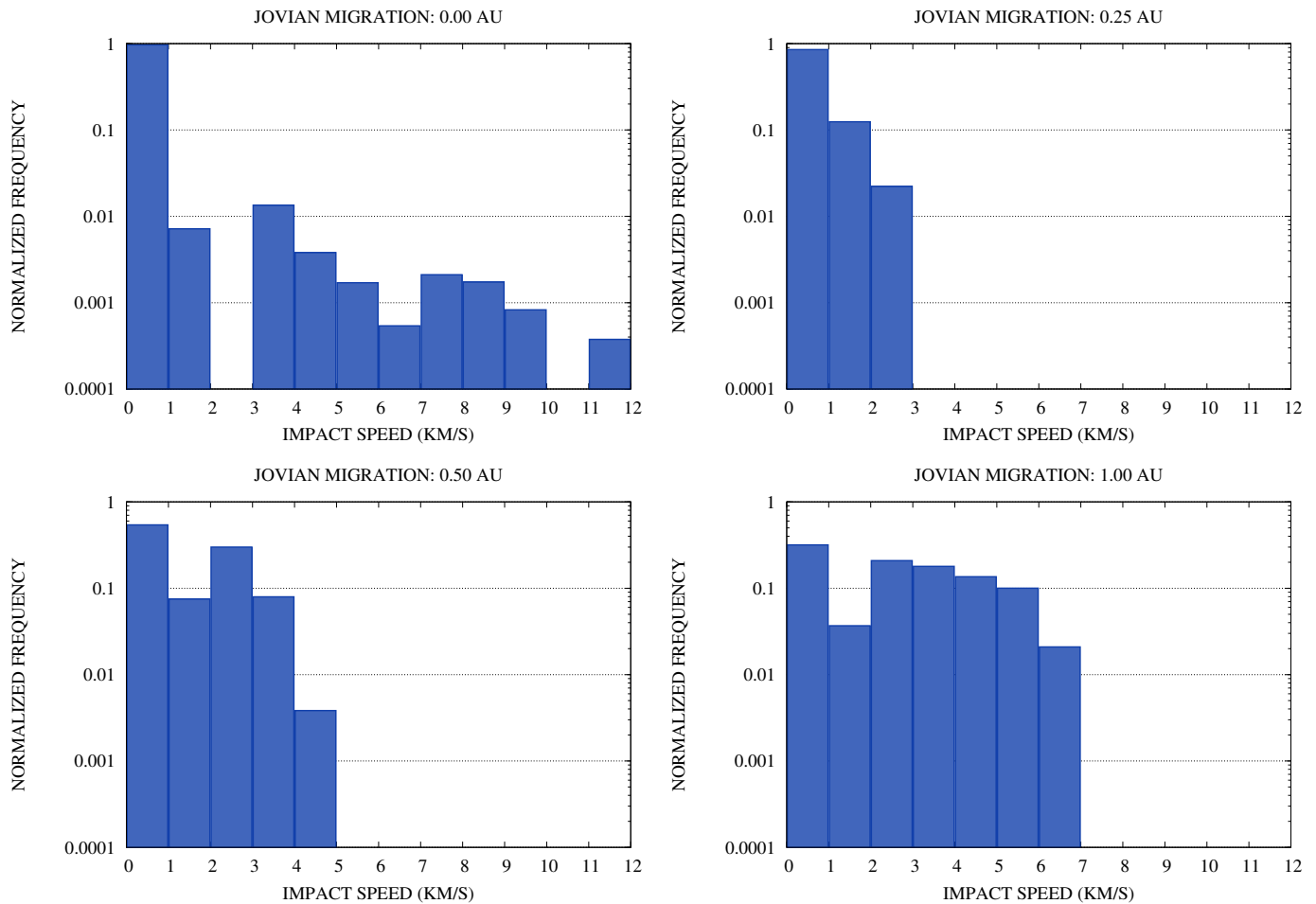}
\caption{Normalized distribution of the impact velocities of ISS impactors reported in Paper I in the four migration scenarios. The distribution of the impact velocities is common to all SFDs of the impactors.}\label{fig1}
\end{figure*}

To better understand the results that will be discussed in the following, it is important to understand what the different SFDs of the planetesimals and Jovian migration scenarios imply in terms of the impact velocities and the SFD of the projectiles hitting Vesta. Fig. \ref{fig1} shows the normalized frequency distributions of the impact velocities, which are common to all SFDs of the planetesimals, in the four migration scenarios. 

In the three scenarios where Jupiter migrates there is a linear correlation between the maximum impact speed and the extent of the Jovian displacement. This is due to the fact that when Jupiter migrates by $0.5$ AU or more also bodies affected by the 2:1 resonance have their eccentricities raised enough to impact Vesta while for Jupiter migrating by $0.25$ AU only the 3:1 resonance produce impactors on the asteroid (see Fig. 1 in Paper I). 

The drop in the impact velocities between the scenario where Jupiter did not migrate and the one where Jupiter migrated by $0.25$ AU is instead due to the fact that when Jupiter migrates part of the planetesimals it excites have their inclinations raised to a few degrees, removing them from the potential impactors on Vesta, which in Paper I was assumed to orbit on the ecliptic. Also this effect increases with increasing displacements of Jupiter 
but it is less important than the appearance of the impactors excited by the 2:1 resonance described previously.


\begin{figure*}[t]
\centering
\includegraphics[width=\textwidth]{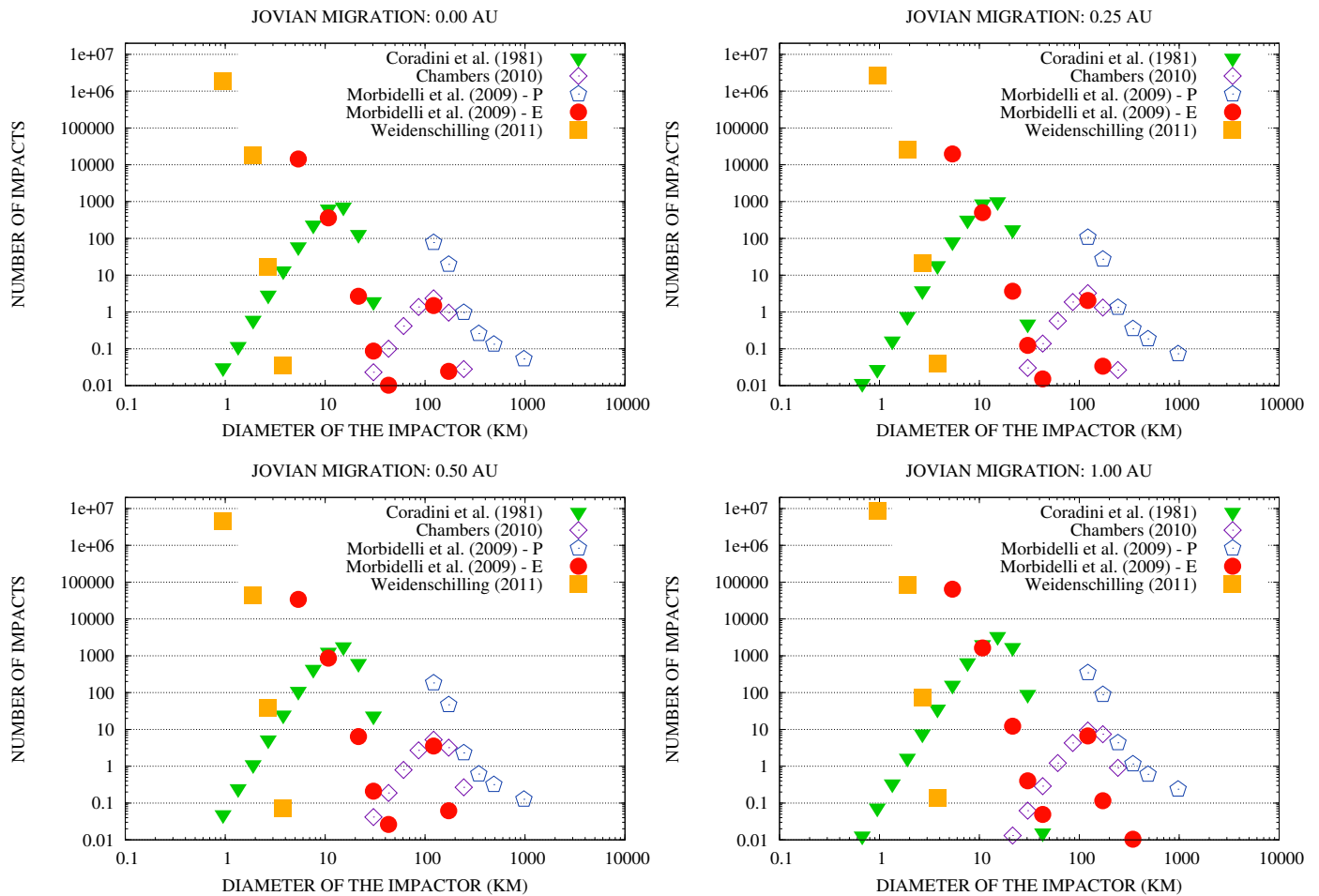}
\caption{Size-frequency distributions of the impactors on Vesta in the four migration scenarios for the different SFDs of the primordial asteroids considered in this work. Numbers of impacts lower than $1$ indicate stochastic events (even when considered cumulatively): these cases are not considered in the following analysis.}\label{fig2}
\end{figure*}

Fig. \ref{fig2} shows instead the size-frequency distributions of the impactors hitting Vesta in the four migration scenarios for the different SFDs of the primordial asteroids considered in this work.
As can be immediately seen, the SFD by \citet{weidenschilling2011} is dominated by $1-2$ km large projectiles, the evolved one by \citet{morbidelli2009} by $5-10$ km large projectiles, the one by \citet{coradini1981} by $10-20$ km large projectiles while the primordial one by \citet{morbidelli2009} and that of \citet{chambers2010} are dominated by $100-200$ km large bodies. Note that those cases where the cumulative number of impacts is lower than $1$ represent stochastic events that are not considered in the following analysis.

As a reference for the values reported in Fig. \ref{fig2}, simulations of the Late Heavy Bombardment performed using the collisional model from Papers I and II and the migration scenario described by \citet{minton2009} indicate that Vesta would be hit by about $300$ impactors with size ranging between $1$ km and $10$ km over a time of $10^{7}$ years (S. Pirani, master thesis at the University of Rome ``La Sapienza'').  

At the end of the simulations of Paper I, all $2\times10^{4}$ ISS test particles were still in the asteroid belt, i.e. they were neither ejected from the Solar System nor collided with Jupiter or the Sun. As discussed in Paper I, test particles that impacted on Vesta became inactive (i.e. they could not impact a second time with the asteroid) but were not removed from the asteroid belt. From a dynamical point of view, therefore, the asteroid belt would not be depleted by the formation of Jupiter alone across the duration of the JEB. From the SFDs of the impactors in Fig. \ref{fig2}, it is possible to verify that the mass removal effect of Vesta is quite limited. In the case of a Jovian migration of $1$ AU, i.e. the most collisionally active scenario among those considered, the mass removed by Vesta through impacts would range from $\sim10^{-6}$ M$_{\oplus}$ (\citealt{coradini1981,weidenschilling2011} and the collisionally evolved case from \citealt{morbidelli2009}) to a maximum of only $\sim10^{-4}$ M$_{\oplus}$ (the primordial case from \citealt{morbidelli2009}).

In this work, therefore, the final mass of the asteroid belt after the JEB is basically the same as the initial one. According to \citet{obrien2007}, once the presence of planetary embryos and of dynamical friction are taken into account, the mass loss from the asteroid belt over the first $1$ Ma following the formation of Jupiter and Saturn should amount to about $10\%$ or, equivalently, the final mass of the asteroid belt should be $90\%$ of the initial one. As the model used in this work assumes that Saturn formed at least after the bulk of the JEB took place, a realistic final mass of the asteroid belt would be between these two extreme value. As a consequence, the likely depletion of the asteroid during the JEB should not affect the order of magnitudes of the effects explored in this work and described in the following.

\subsection{Survival of Vesta}

\begin{table*}[t]
\centering
\begin{tabular}{cccccc}
\hline
 \textbf{SFD of planetesimals} & & \textbf{Migration} &  \textbf{scenario} & \T\B \\
\hline
 & $0.00$ AU & $0.25$ AU & $0.50$ AU & $1.00$ AU \T\B\\
\hline
 \citet{coradini1981} & $0$ & $0$ & $0$ & $0$ \T \\
 \citet{chambers2010} & $<10^{-2}$ & $<10^{-2}$ & $<10^{-2}$ & $1.55\times10^{-2}$ \T \\
 \citet{morbidelli2009} - P & $1.98\times10^{-2}$ & $<10^{-2}$ & $0.20$ & $0.91$  \T \\ 
 \citet{morbidelli2009} - E & $0$ & $<10^{-2}$ & $<10^{-2}$ & $<10^{-2}$ \T \\ 
 \citet{weidenschilling2011} & $0$ & $0$ & $<10^{-2}$ & $<10^{-2}$  \T\B\\
\hline
\end{tabular}
\caption{Number of catastrophic impacts on Vesta across the JEB in the different migration scenarios and for the different SFDs considered. The labels ``E'' and ``P'' of the SFDs from \citet{morbidelli2009} identify respectively the collisionally evolved one and the primordial one.}\label{table1}
\end{table*}

Table \ref{table1} shows the number or, more properly, the probability of Vesta undergoing a catastrophic collision across the JEB, averaged over $10^{4}$ Monte Carlo extractions. These refined results are in agreement with those of \citet{bottke2005a,bottke2005b} and confirm the findings of Paper I: Vesta, and more generally planetesimals of analogous size (see Paper II), have little chances of being disrupted by catastrophic impacts caused by the JEB. As pointed out in Paper II, their fate across the JEB is determined instead by cratering erosion. Note that the shattering effects of high-energy impacts are not considered as, at the time of the JEB, Vesta consisted of a molten interior topped by a solid crust of thickness no larger than $30$-$40$ km \citep{formisano2013,schiller2011}. Fractures and cracks produced by the JEB in the solid crust were likely filled by magmatic intrusions that, once solidified, counteracted the shattering effects of the most energetic impacts. Depending on the considered scenario for the evolution of the crust (respectively, that of \citealt{formisano2013,tkalcec2013} and that of \citealt{schiller2011}), the material filling these fractures could be either a mixture of molten eucrites and diogenites or olivine-rich material from the mantle. 

\subsection{Crater saturation of the Vestan surface}

To assess the degree of saturation of the Vestan surface after the JEB, this study computed the R-value distributions of the crater populations produced by the different SFDs of the primordial asteroids considered. 
This study took as the smallest crater diameter the value of $0.1$ km and, following \citet{melosh1989}, divided the crater population associated to each SFD in bins where $D_{i+1}=\sqrt{2}D_{i}$ and the central diameter  is the geometric mean $\overline{D}=\sqrt{D_{i}D_{i+1}}$, and computed the R-value as
\begin{equation}
R_{i}=3.65 f_{i}
\end{equation} 
where $f_{i}$ is the fraction of the Vestan surface covered by the craters in the relevant bin. The value of $f_{i}$ of each bin is obtained simply by summing the surface areas $A$ covered by the $N$ craters in the bin and dividing it by the surface of Vesta $S_{V}$:
\begin{equation}
f_{i}=\left( \Sigma_{j=0}^{N_{i}} A_{j} \right)/S_{V}
\end{equation} 
where $A_{j}$ is the geometrical area of each crater of diameter $D_{j}$:
\begin{equation}
A_{j}=\frac{\pi}{4}D_{j}^2
\end{equation}
Note that, when the condition $D_{i+1}=\sqrt{2}D_{i}$ is satisfied and the average diameter of each bin is computed as the geometric mean $\overline{D}=\sqrt{D_{i}D_{i+1}}$, this definition of the R-value is equivalent to the one given by the \citet{crater1978}, i.e.
\begin{equation}
R_{i}=\frac{N_{i}\,\overline{D}^{3}}{S_{v}\left(D_{i+1}-D_{i}\right)}
\end{equation}
In building the R-value distributions, only the effects of low-energy impacts ($Q < 0.1Q^{*}_{D}$) were considered: high-energy and catastrophic impacts were not included. Moreover, this study considered only those bins where the cumulative impact probability produced at least $1$ impact once normalized to the real population of the disk.
 
As a comparison, this study also computed the R-value distribution produced by the last $3.5$ Ga of collisional evolution of Vesta. To do this, this study based on the method and the assumptions made by \citet{mccord2012} in estimating the flux of low-albedo impactors on Vesta. This estimate was based on the SFD of the present asteroid belt described by \citet{bottke2005a}, corrected for the depletion of a factor $2$ suggested to have occurred over the last $3.5$ Ga by \citet{minton2010}. The intrinsic impact probability and the average impact speed computed by \citet{obrien2011} for Vesta were then used to estimate the SFD of the craters produced on Vesta over the considered time interval.

\begin{figure*}[t]
 \centering
 \includegraphics[width=\textwidth]{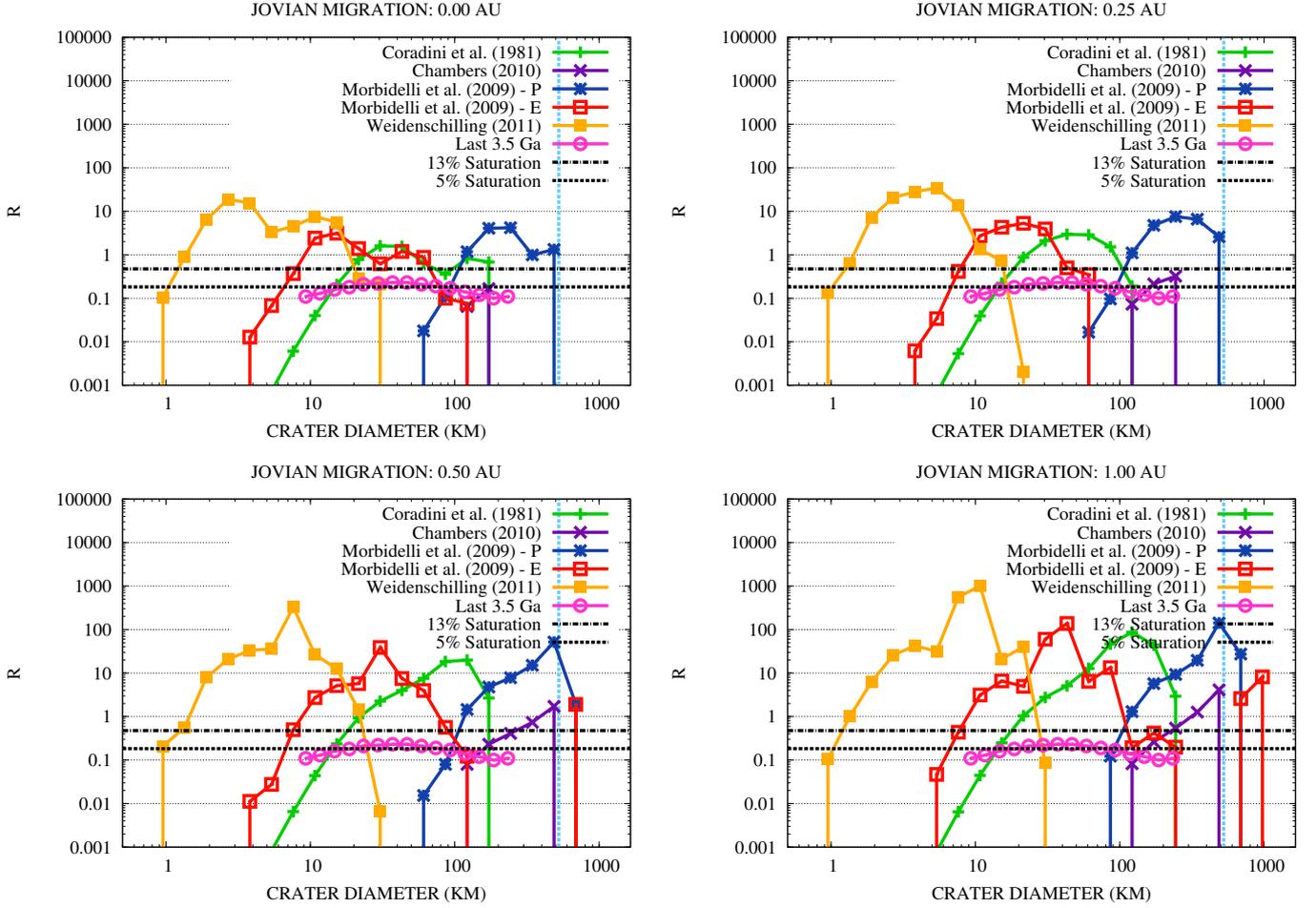}
 \caption{R-plot of the crater populations on Vesta after the JEB for the different SFDs considered in this work. The $5\%$ and $13\%$ saturation levels and the crater population expected due to the last $3.5$ Ga of collisional evolution of Vesta are shown for reference. The contribution of cometary impactors from Paper I is not shown here. The labels ``E'' and ``P'' of the SFDs from \citet{morbidelli2009} identify respectively the collisionally evolved one and the primordial one. The light blue vertical dashed line indicates the present diameter of Vesta.}\label{fig3}
\end{figure*}

The results obtained for the different SFDs and Jovian migration scenarios are shown in Fig. \ref{fig3}, where the $5\%$ and $13\%$ saturation levels were also indicated. These two threshold levels represent respectively the minimum R-value for which a crater population can reach equilibrium and the R-value estimated for Mimas, whose surface is the most densely cratered in the Solar System \citep{melosh1989}. As can be identified at first glance, the population of craters produced by the JEB across $1$ Ma is significantly larger than the one produced over the last $3.5$ Ga, often by orders of magnitude. 
All the considered SFDs of impactors caused the saturation of the surface of Vesta at levels higher than $13\%$. This is not surprising if we consider that the original population of the asteroid belt was likely $100-1000$ times larger than the present one \citep{weidenschilling1977,weidenschilling2011,morbidelli2009} and that the volume they occupied was $10-100$ times smaller due to their lower orbital inclination values (e.g. Paper I assumed an initial maximum inclination of $\sim1.7\degree$ while the present value is $\sim30\degree$).

The populations of craters produced by the SFDs from \citet{coradini1981,morbidelli2009,weidenschilling2011} span different ranges of diameters but are overall similar. The SFD by \citet{coradini1981} produces a crater population ranging between $10$ km and $250$ km. The SFD by \citet{weidenschilling2011} produces instead a crater population ranging between $1$ km and $30$ km. Due to the higher abundance of small impactors respect to the other SFDs, no large ($D > 100$ km) craters are produced during the JEB in this case.  Finally, the collisionally evolved SFD by \citet{morbidelli2009} produces an intermediate case between the previous two, with the crater population ranging between $4$ km and $200$ km. With the latter SFD, however, in the cases of Jupiter migrating by $0.5$ AU and $1$ AU Vesta would undergo respectively $1$ and $3$ high-speed impacts with bodies with diameter of about $100$ km, which would produce craters with diameters larger than the diameter of Vesta (see the points beyond the vertical dashed line in Fig. \ref{fig3}). These cases are therefore incompatible with the survival of the basaltic surface of Vesta. Finally, the SFD by \citet{chambers2010} and the primordial one by \citet{morbidelli2009} represent a separate class of results, as they only produce craters in the range $60-500$ km in diameter. Also in the case of the primordial SFD of \citet{morbidelli2009} Vesta would undergo impacts whose craters would be larger than its diameter if Jupiter migrated by $0.5$ AU or more (see the points beyond the vertical dashed line in Fig. \ref{fig3}).

\subsection{Mass loss of Vesta}

\begin{figure}[t]
 \centering
 \includegraphics[width=3.6in]{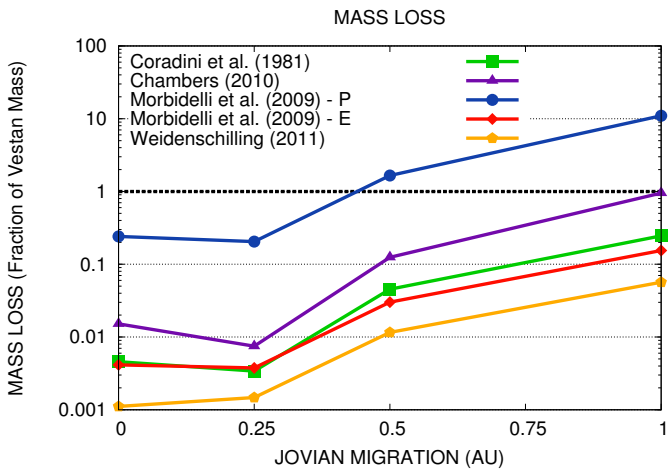}
 \caption{Mass loss of Vesta during the JEB in the different migration scenarios and for the different SFDs considered. Mass loss is expressed in units of the present Vestan mass. The dashed line indicates the present mass of Vesta according to \citet{russell2012,russell2013}.}\label{fig4}
\end{figure}

Eqs. \ref{erosion_svetsov} and \ref{erosion_benz} allow to compute the mass loss of Vesta due to the different combinations of Jovian migration and SFD of the impactors. The results are shown in Fig. \ref{fig4}, where they are expressed in units of the present mass of Vesta. 

In the olivine-rich case discussed by \citet{ruzicka1997}, where the radius of Vesta was set to $R_{V}=265$ km, the eucritic layer would account for about $26.6\%$ of the volume of Vesta while the diogenitic layer would account for about $11.3\%$ of the volume of the asteroid. Assuming a density value for eucrites of $3000$ kg m$^3$ \citep{consolmagno1998} and expressing the results in units of the present mass of Vesta, the mass values corresponding to these volumes would be $24\%$ for the eucritic layer and $10\%$ for the diogenitic layer.

The SFD by \citet{weidenschilling2011} is the least erosive one among those considered. For Jupiter migrating by $0.25$ AU or less, the eroded mass ranges between $0.1\%$ and $0.15\%$. The mass erosion grows to $1\%$ when Jupiter migrates by $0.5$ AU and jumps to $6\%$ if Jupiter migrates by $1$ AU.

The collisionally evolved SFD by \citet{morbidelli2009} and the one by \citet{coradini1981} produce very similar values of the mass loss of Vesta. For Jupiter migrating by $0.25$ AU or less, the eroded mass is about $0.5\%$. The mass erosion grows respectively to $3-4.5\%$ when Jupiter migrates by $0.5$ AU and to $15-25\%$ when Jupiter migrates by $1$ AU. 

The SFD by \citet{chambers2010} is compatible with the survival of Vesta only in those scenarios where the Jovian migration is limited ($0.25$ AU or less). For a Jovian displacement of $0.5$ AU, Vesta would lose $12\%$ of its mass, while for a Jovian displacement of $1$ AU the asteroid would not survive the JEB. Note, however, that the inclusion of OSS impactors significantly increases the number of high-energy impactors in all migration scenarios and, as shown by Paper II, generally makes this SFD inconsistent with the survival of Vesta. 

Finally, the case of the primordial SFD by \citet{morbidelli2009} is even more drastic than the one by \citet{chambers2010}, as Vesta (or its basaltic crust) would not survive the JEB independently on the extent of the migration of Jupiter and of the inclusion of the OSS impactors.

As a comparison, using Eq. \ref{erosion_svetsov} together with the decay law of the asteroid belt population over the last $3.98$ Ga \citep{minton2010} and the model described by \citet{mccord2012}, and considering all possible impactors, the secular erosion of Vesta  over the last $3.98$ Ga can be estimated to range between $2.7\%$ (if the surface of Vesta can still be modelled as rock at the time) and $0.28\%$ (if instead its behaviour is that of regolith) of the present mass of the asteroid.

\subsection{Surface erosion of Vesta}

To better understand the implication of the JEB for the survival of the basaltic crust of Vesta, the mass loss values described above can be expressed in terms of the thickness $\Delta R$ of the shell, starting from the present surface of Vesta and extending outward, whose mass matches the mass $\Delta M$ lost by the asteroid, i.e.:
\begin{equation}
\Delta R = \left( R_{V}^{3} + \frac{3\Delta M}{4\pi\rho_{V}} \right)^{1/3} - R_{V}
\end{equation}
where $R_{V}=262.7$ km and $\rho_{V}=3090$ kg m$^{-3}$ are respectively the mean radius and the mean crustal density of Vesta measured by the Dawn mission \citep{russell2012,russell2013}. Before discussing the surface erosion of Vesta, it should be stressed that the values here computed for the thickness of the eroded layer cannot be compared directly with the thickness of the eucritic and diogenitic layers estimated by \citet{ruzicka1997}. As mentioned previously, in fact,  \citet{ruzicka1997} assumed that Vesta originally had a mean radius of $265$ km, so a direct comparison is possible only in cases where the surface erosion is of about $2-3$ km.  

\begin{figure*}[t]
 \centering
 \includegraphics[width=\textwidth]{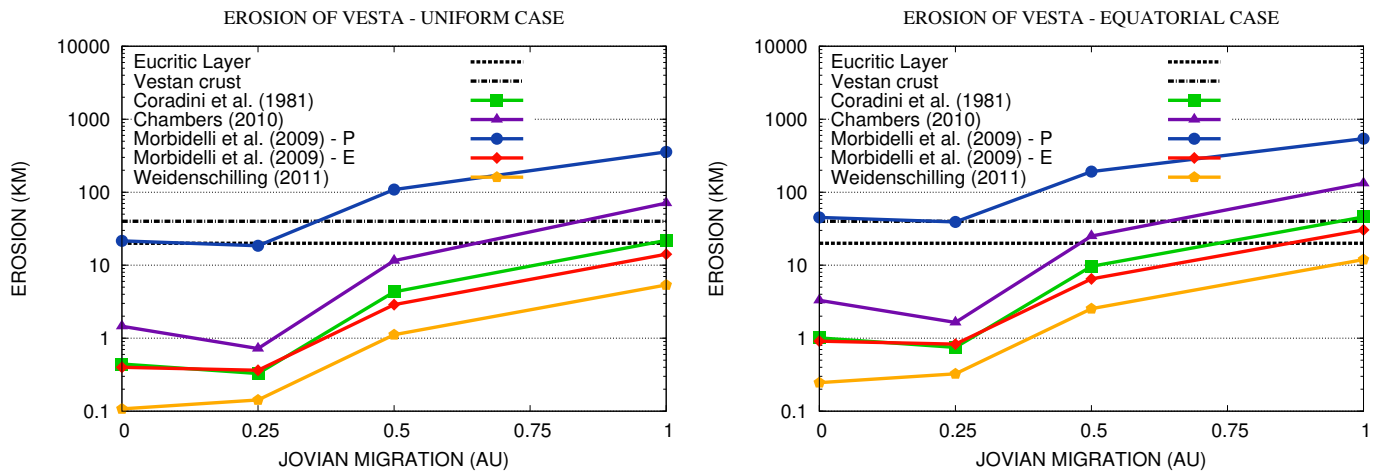}
 \caption{Surface erosion of Vesta during the JEB in the different migration scenarios and for the different SFDs considered. Surface erosion is expressed as the thickness of a shell with density equal to the present average one of Vesta and mass equal to the mass loss, extending from the present radius of Vesta outwards. The plot on the left shows the case of a uniform erosion, the plot on the right shows instead the case of erosion proportional to the cross-sectional area of Vesta during the JEB. The labels ``E'' and ``P'' of the SFDs from \citet{morbidelli2009} identify respectively the collisionally evolved one and the primordial one. The dashed and dot-dashed lines indicate respectively the thickness of the eucritic layer and of the whole vestan crust according to \citet{ruzicka1997}.}\label{fig5}
\end{figure*}

Fig. \ref{fig5} shows the degree of surface erosion of Vesta in the considered cases under two different assumptions on the distribution of the impacts on the surface of the asteroid. While shown in the plots of Fig. \ref{fig5} for reasons of completeness, the primordial SFD by \citet{morbidelli2009} is never compatible with the survival of Vesta or of its basaltic crust. As a consequence, it will not be discussed in the following.

The left-hand plot of Fig. \ref{fig5} shows the degree of surface erosion estimated by assuming that Vesta lost mass uniformly from all its surface. For the SFD supplied by \citet{weidenschilling2011}, in the cases of limited ($0.25$ AU) or no migration of Jupiter the pre-JEB Vesta needed to be $100-150$ m larger to account for the lost mass. The thickness of the eroded layer jumps to about $1$ km if Jupiter migrated by $0.5$ AU and to $5$ km when the Jovian displacement is of $1$ AU.

In the cases of the SFD by \citet{coradini1981} and the collisionally evolved one from \citet{morbidelli2009}, the thickness of the eroded layer is about $300-500$ km if Jupiter's migration was limited ($0-0.25$ AU). The thickness of the eroded layer grows to $3-4$ km for Jupiter migrating by $0.5$ AU and to about $14-22$ km for a migration of $1$ AU. 

As previously mentioned, for the SFD by \citet{chambers2010} the only cases possibly compatible with the survival of the basaltic crust of Vesta are those where Jupiter migrates by $0.25$ or less: the thickness of the eroded layer then is of about $1$ km. As soon as  larger displacements of Jupiter as considered, the thickness of the eroded layer rapidly goes to tens of km. 

Given the more compact configuration of the planetesimals at the time of Jupiter's formation and of the JEB, it is possible that impacts on Vesta had a preferential direction instead of being isotropic. If impacts took place mostly on the orbital plane of Vesta (which, in the simulations of Paper I, was the same as the midplane of the circumsolar disk), the distribution of the impacts would be a function of the cross-sectional area of the asteroid. Assuming that the primordial Vesta was spherical, simple geometric arguments show that the region between $\pm45\degree$ of latitude accounts for about $70\%$ of the cross-sectional area of the asteroid. As a consequence, about $70\%$ of the impacts during the JEB should hit Vesta in this ``equatorial'' belt. The remaining $30\%$ of the impacts should hit Vesta in the two ``polar'' regions going from $+45\degree$ to $+90\degree$ and from $-45\degree$ to $-90\degree$ of latitude. 

As shown in the right-hand plot of Fig. \ref{fig5}, in this case the thickness of the excavated layers across the ``equatorial'' belt is about twice as big as the corresponding values reported in the case of uniform erosion. As a consequence, the SFD by \citet{chambers2010} would remove $2-3$ km from the vestan crust in the previously discussed cases. The collisionally evolved SFD by \citet{morbidelli2009} and the one by \citet{coradini1981} are compatible with the survival of the crust of Vesta only for Jovian displacements smaller than $0.5$ AU: in these cases, the thickness of the eroded layer is about $1$ km. The case of $0.5$ is borderline, with a crustal erosion of $7-10$ km.  The SFD by \citet{weidenschilling2011} is the most favourable for the survival of the crust of Vesta. For increasing values of the Jovian displacement, the thickness of the eroded layer goes from $200-300$ m to $2.5$ km and then to $12$ km.

\subsection{Excavation of the mantle}\label{excavation}

Given that in most cases the JEB can saturate the surface of Vesta up to crater diameters in the range of $100-200$ km (see Fig. \ref{fig3}), this study also assessed the number of craters that can excavate the whole basaltic crust of Vesta and extract olivine-rich material (possibly crystallized but mostly likely molten, see e.g. \citealt{formisano2013} and \citealt{tkalcec2013}) from the mantle of the asteroid. Then, the lack of olivine in the bulk of the HED family of meteorites \citep{mcsween2011} and of olivine signatures in the spectra collected from Dawn \citep{desanctis2012a} can be used to rule out unrealistic cases. 

\begin{figure*}[t]
 \centering
 \includegraphics[width=\textwidth]{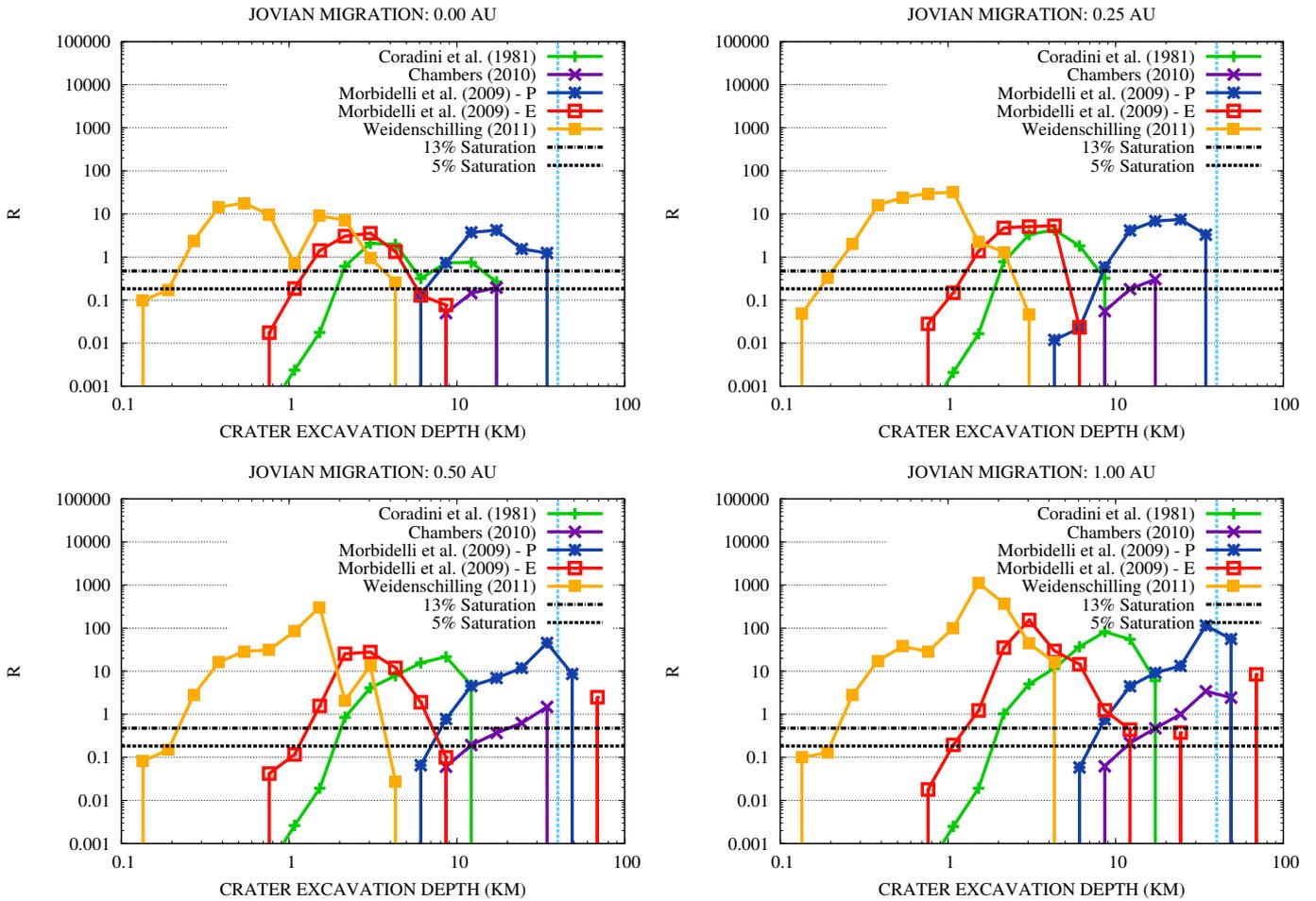}
 \caption{R-plot of the excavation depth of craters on Vesta across the JEB for the different SFDs and migration scenarios considered in this work. The crater depths are derived from Fig. \ref{fig3} assuming a depth-to-diameter ratio of $0.168$ (Vincent et al., this issue). The $5\%$ and $13\%$ saturation levels are shown for reference. The contribution of cometary impactors from Paper I is not shown here. The labels ``E'' and ``P'' of the SFDs from \citet{morbidelli2009} identify respectively the collisionally evolved one and the primordial one. The vertical light blue dashed line indicates the thickness of the crust of Vesta (i.e. the eucritic and diogenitic layers) as estimated by \citet{ruzicka1997}.}\label{fig6}
\end{figure*}

According to Vincent et al. (this issue), on Vesta the transition from simple to complex craters seems to occur at diameters of about $30$ km. For craters smaller than this value a constant depth-to-diameter ratio of $0.168$, i.e. equal to the average value measured on Vesta by Dawn (Vincent et al., this issue), was assumed. For larger craters the conservative relation from \citet{melosh1989}
\begin{equation}\label{crater_depth}
d_{exc}=0.1\,D_{t}=0.1\left(D/1.3\right)=0.077\,D
\end{equation}
was used, where $d_{exc}$ is the depth of excavation, $D_{t}$ is the diameter of the transient crater and $D$ is the final diameter of the crater. The factor $1.3$ is used to scale the diameter of the transient crater to that of the final crater \citep{holsapple1993}. Using the results of Fig. \ref{fig3}, the excavation of Vesta can then be expressed in the form of an R-plot of the crater depths across the JEB, as shown in Fig. \ref{fig6}.

It is immediately evident that the SFD from \citet{chambers2010}, when Jupiter migrates by $0.5$ AU or more, and the primordial SFD from \citet{morbidelli2009} saturate the surface of Vesta with craters that excavate Vesta at depths ranging between $30$ km and $50$ km (see Fig. \ref{fig6}). As a consequence, significant quantities of olivine-rich material should be excavated across the JEB and mixed with the regolith. In the cases of limited (i.e. up to $0.25$ AU) and no migration of Jupiter, the SFD from \citet{chambers2010} would saturate at least at $5\%$ level the surface of Vesta with craters capable of excavating the whole eucritic layer and exposing or extracting material from the diogenitic one.

The case of the SFD by \citet{weidenschilling2011} is opposite to the previous ones (see Fig. \ref{fig6}): the crater population associated to this SFD only excavates as deep as the eucritic layer and never reaches the diogenitic lower crust nor the olivine-rich mantle. The case of the collisionally evolved SFD from \citet{morbidelli2009} is similar when the Jovian migration is limited (i.e. up to $0.25$ AU) or absent. In the cases where Jupiter migrated by $0.5$ AU and $1$ AU Vesta would undergo respectively $1$ and $3$ high-velocity impacts with bodies of the order of $100$ km in diameter (see the isolated red points beyond the vertical dashed line in the bottom panels of Fig. \ref{fig6}). As observed in discussing Fig. \ref{fig3}, these events would destroy the basaltic crust of Vesta and expose the olivine-rich mantle on an hemispheric level.

Finally, the case of the SFD by \citet{coradini1981} is similar to that of \citet{weidenschilling2011} when the migration of Jupiter is moderate (i.e. between $0.25$ AU and $0.5$ AU), as the crater population mainly excavates and reprocesses the eucritic layer. If the giant planet migrated by $1$ AU or did not migrate, the JEB would saturate the vestan crust with craters capable of excavating the whole eucritic layer and exposing or extracting material from the diogenitic one. 

\subsection{Formation of a regolith layer and effusive phenomena}\label{regolith}

While a detailed assessment of the regolith production during the JEB is beyond the purpose of this paper, a simple estimate can be easily made. Paper I originally estimated the erosion of Vesta by simply summing the volumes of the material excavated to form the craters (see Tables 3, ``post-core'' column, and 5 from Paper I), neglecting the effects of the relatively high escape velocity of the asteroid. By comparing the values estimated by these authors for the thickness of the eroded layer with the corresponding values reported in Fig. \ref{fig5}, we can easily see that only a fraction of the excavated material is effectively lost to Vesta.

This implies that the ejecta produced by the JEB should form a layer of regolith and mega-regolith whose thickness can easily range between $1$ km to a few km. It may also imply that, across the bombardment, the mechanical properties of the surface of Vesta should change from those of solid rock to those of regolith. This change would result, in turn, in a corresponding change in the scaling laws governing the formation of craters, at least for those impacts that do not excavate deeper than the regolith layer itself, and in cratering erosion rates about an order of magnitude lower (see e.g. \citealt{holsapple2007}). 

However, the surface evolution of Vesta depends strongly on the geophysical state of the asteroid. As discussed by \citet{formisano2013}, depending on the accretion time and the initial porosity, the thickness of the solid crust of Vesta across the JEB could have ranged from a minimum of $7-9$ km to a maximum of $20-30$ km. The independent study by \citet{tkalcec2013}, using a more complete physical model, gives thickness values for the solid crust varying between $7$ km and at least $10$ km (\citealt{tkalcec2013}, Supplementary Information; G. Golabek, personal communication) between $3$ Ma and $5$ Ma from CAIs, and up to $20-30$ km at about $9$ Ma \citep{tkalcec2013}.

For these values of the thickness of the solid crust of Vesta, it is plausible that the JEB caused effusive phenomena on a regional or global scale on the asteroid, as originally proposed by Paper I. The molten material would quickly crystallize due to radiative effects once brought to the surface and would therefore re-compact the regolith and counteract the effects of the ejecta blanketing.


\section{Discussion}

The current work investigated the surface evolution and the erosion of Vesta during the JEB using a more refined collisional model and more detailed calculations than those of Papers I and II. The assessment of the effects of the JEB using this improved model, however, is still based on the original simulations of Paper I: as such, the effects of gas drag and the perturbations of the planetary embryos are omitted in the dynamical model of the Solar Nebula. The implications of these neglected processes for the results of this work will be discussed later.

\subsection*{Catastrophic disruption and crater saturation}

The first result of this work is the confirmation that, across the JEB, the chances of Vesta undergoing a catastrophic collision with another planetesimal are always negligible, as originally noted by Paper I and in agreement with the findings of \citet{bottke2005a,bottke2005b}. Only in the most extreme scenarios, which are however deemed unrealistic, the chances can increase further than $1\%-2\%$. 

In all investigated scenarios, the JEB saturates the surface of the asteroid with craters. The Dawn mission recently revealed that a large fraction of the surface of Vesta has been reset by large impacts and does not preserve a record of the most ancient collisional activity of the asteroid \citep{marchi2012,russell2013}. The regions showing the highest crater density are possibly saturated and a few geologic features on the surface of Vesta have been tentatively identified as more ancient, degraded large craters \citep{marchi2012}.

It is implausible that the observed saturation and degraded features can be related to the JEB. By extrapolating from the number of craters produced across the last $3.5$ Ga (see Fig. \ref{fig3}) and taking into account the effects of the Late Heavy Bombardment (S. Pirani, master thesis at the University of Rome ``La Sapienza''), an educated guess is that the cratering record  on Vesta should not allow us to probe much earlier than $\sim4$ Ga ago. Should the latter estimate prove correct, the first few $10^{8}$ years of the life of Vesta (and of the Solar System) would be precluded to us in terms of crater record. 

Catastrophic impacts and the crater record of Vesta would therefore appear unable to supply more details on the primordial history of the asteroid respect to what is already known.


\subsection*{Planetesimals and turbulent disks}

The survival of Vesta across the JEB is generally not compatible with impactors in the size range predicted for planetesimals that formed in turbulent disks (i.e. the SFD from \citealt{chambers2010} and the primordial SFD from \citealt{morbidelli2009}, see Fig. \ref{fig2}). Cratering erosion would strip the asteroid of a significant fraction of its mass, causing the loss of half or all the eucritic layer, while large impacts would form basins capable to excavate mantle material on a global scale, in contrast with available meteoritic and spectral data.

The only exception to this picture are the cases of the SFD by \citet{chambers2010} and of limited (i.e. less than $0.25$ AU) or no migration, where the mass loss and the surface erosion of Vesta estimated in this work would be in principle compatible with the survival of the vestan crust. Even in these cases, however, the vestan surface would be saturated at least to a $5\%$ level by craters capable of exposing the diogenitic layer, incompatibly with its present day limited exposure.
Moreover, the results of Papers I and II indicate that the contribution of OSS impactors, here not considered, would significantly increase the number of basin-forming impacts, and consequently the mass loss of Vesta, to a level inconsistent with the survival of the vestan crust. 

\citet{morbidelli2009} assumed that Jupiter formed $3$ Ma after CAIs based on the median lifetime of circumstellar disks and showed that, by that time, the SFD of the asteroid belt would have evolved into a shape more similar to its present one, though scaled to a larger population of asteroids. If a turbulent nebula was the birthplace of the primordial planetesimals, these results therefore imply that Jupiter should have formed no earlier than the time it requires for the asteroid belt to reach the final state described by \citet{morbidelli2009}, if this time is shorter than the assumed $3$ Ma, for Vesta to survive the JEB.

In this case, however, the results of the collisional model rule out large ($0.5$ AU or larger) displacements of Jupiter. The collisionally evolved SFD by \citet{morbidelli2009} is compatible with observational data and with the formation of Rheasilvia and Veneneia without exposing the olivine-rich mantle only in the cases of no migration and $0.25$ AU migration of the giant planet, as the crustal erosion is limited to about $1$ km. If Jupiter migrated $0.5$ AU or more, the collisionally evolved SFD by \citet{morbidelli2009} would imply between $1$ and $3$ basin-forming impacts on Vesta, capable of destroying the crust and exposing the mantle on a hemispheric scale.

\subsection*{Planetesimals and quiescent disks}

The SFDs of the primordial asteroids that are based on the assumption that planetesimals formed in a quiescent disk are more favourable to the survival of Vesta than those where a turbulent nebula is assumed.

The SFD by \citet{coradini1981} is generally characterized by mass loss and erosion rates slightly larger than those of the collisionally evolved SFD by \citet{morbidelli2009}. As such, the most likely scenario appears the one of Jupiter undergoing a limited (i.e. about $0.25$ AU) migration. If Jupiter migrated by $1$ AU the eucritic layer and a large part of the diogenitic one would be stripped off by the JEB. If the Jovian migration was less extensive (i.e. $0.5$ AU), Vesta would still lose a significant part ($\sim5$ km) of its eucritic layer, making its survival to the Rheasilvia-forming and Venenenia-forming impacts less likely. In the case of no migration of Jupiter, the vestan surface would be saturated a $5\%$ level by craters capable of exposing the diogenitic layer, incompatibly with its present day limited exposure.



The SFD by \citet{weidenschilling2011} is the one that poses the least constraints to the migration of Jupiter and best fits the observational constraints on Vesta. It is the most compatible with the survival of the basaltic crust of the asteroid, as in the case of no migration and for Jovian displacements of $0.25$ AU and $0.5$ AU the surface erosion never exceeds $1$ km. No large basins are formed across the JEB and the excavation due to impact cratering is always limited to the eucritic layer. Following \citet{morbidelli2009}, also \citet{weidenschilling2011} assumed a formation time of Jupiter of $3$ Ma from CAIs but noted that, in general, most of the essential features of the asteroid belt are already achieved after $1$ Ma. In this case, the results previously described should in principle hold true also if Jupiter formed between $1$ and $3$ Ma after CAIs. \citet{weidenschilling2011} also pointed out that the collisional evolution of the asteroid belt proceeds at a lower rate after the first $1$ Ma. Given its short duration ($0.5$-$1$ Ma), the assumption of a fixed SFD during the JEB then should not affect the results in a significant way.

\subsection*{Surface evolution and regolith production}

A side effect of the JEB in all explored scenarios is the production of a large amount of ejecta excavated by the impacts that saturate the surface of the asteroid. The largest part of these ejecta does not escape the relatively strong gravity of Vesta and fall back on the surface. This causes the creation of a regolith layer that can easily be a few km thick, as can be seen by comparing the erosion rates in this work with those from Paper I. The appearance of a thick regolith layer, in turn, can affect the thermal evolution of the asteroid, by more efficiently insulating its interior, and the outcomes of the cratering processes, by modifying the erosion rate and the size distribution of at least the smaller craters. The erosion rate, in particular, could be lowered by about an order of magnitude, making it more easy for Vesta to preserve its basaltic crust. 

It must be noted that, strictly speaking, the previous results and their interpretation apply only to the standard scenario where the eucritic and diogenitic layers already solidified at the time the JEB occurred (\citealt{bizzarro2005,schiller2011}, see also \citealt{coradini2011}). The results of \citet{formisano2013} and \citet{tkalcec2013} in modelling the differentiation of Vesta, however, indicate that the diogenitic layer and part of the eucritic layer likely were in a partially molten state at the time of the JEB. This scenario is still compatible with the meteoritic data on HEDs \citep{bizzarro2005,schiller2011}, but the formation process of regolith (and megaregolith) on Vesta would be counteracted by impact-triggered effusive phenomena and by magmatic intrusions into the fractures in the solid crust, as they would recompact it once solidified. Detailed studies of the coupling between the geophysical and the collisional evolution of Vesta are therefore required to fully address the primordial history of the asteroid.

\subsection*{The JEB and the undifferentiated primordial crust}

Another important result of thermal modelling that affects directly the results of this work and their interpretation is the finding by \citet{formisano2013} that an unmelted layer (likely of chondritic composition) would survive the differentiation of Vesta and overlie the eucritic layer. Even if not explicitly stated, this finding has been independently confirmed (G. Golabek, personal communication) by the results of the simulations by \citet{tkalcec2013}. According to both groups, the thickness of this unmelted layer would be at least of about $3$ km (see \citealt{formisano2013}). 

In this case, the previous discussions of the most plausible Jovian migration scenarios for the different SFDs would not be valid. On the contrary, the most plausible scenarios would be the ones where the uniform erosion of the solid crust of Vesta is of the order of one km or larger, as the undifferentiated crust has to be removed to produce the present-day Vesta. The results of this study would therefore favour displacements of Jupiter of about $0.5$ AU for the SFD by \citet{coradini1981} and the collisionally evolved one by \citet{morbidelli2009} (but the latter would still imply $1$ impact capable of exposing the mantle of Vesta) and of $0.5$ AU or more for the SFDs from \citet{weidenschilling2011}.

\subsection*{Limitations of the model and sources of uncertainty}

Before concluding, it is important to point out that the results of this work should be regarded only as a first exploration of the mostly unknown primordial collisional history of Vesta. Poorly constrained factors, as the formation region of the asteroid, can significantly change the global picture. Since this work in based on the simulations performed by Paper I, Vesta is assumed to have formed near its present orbit. This is a reasonable assumption but is not the only possibility. While the results of Paper II rule out the possibility that Vesta formed farther away from the Sun, it is possible that Vesta formed on a inner orbit and was scattered outward during the phase of excitation and depletion of the asteroid belt (see \citealt{coradini2011,obrien2011} and references therein). A inner orbit located farther away from the orbital resonances with Jupiter can result in a lower flux of impactors on Vesta (see also \citealt{weidenschilling1975}) and, therefore, in a milder JEB.

An important source of uncertainty lies in the modelling of the Solar Nebula in the simulations of Paper I. While the interested readers are referred to Papers I and II for more details on the limitations of the dynamical model, a short summary will be presented here to give a clearer picture of their implications for the results of this work. As mentioned previously, neither the effects of gas drag nor the gravitational perturbations of planetary embryos were accounted for in the simulations of Paper I. Planetary embryos would excite the planetesimals, thus rising the impact velocities, and would inject new planetesimals into the Jovian resonances, thus increasing the flux of impactors. On the other hand, the combined effects of the Jovian resonances and planetary embryos would trigger the depletion of the asteroid belt and reduce the population of planetesimals already during the JEB. Due to the short duration of the JEB, the latter effect should not significantly affect the results presented here as the depletion should amount to about $10\%$ at most (see \citealt{obrien2007} and  Papers I and II for a discussion). 
Gas drag would damp the eccentricities of the planetesimals, thus reducing their relative velocities, and would impose them a radial inward migration that would bring new bodies inside the Jovian resonances (for a discussion of the latter see \citealt{weidenschilling2001}). The interplay between the excitation caused by planetary embryos and the damping due to gas drag could have important implications for the JEB, especially in the case of the SFD by \citet{weidenschilling2011} where the impactors are characterized by small sizes. 

Finally, another source of uncertainty is linked to the assumed description of the formation and dynamical evolution of Jupiter.  For what it concerns the latter, Paper I tested what would be the implications of a slower migration of Jupiter for the JEB and found that it should not affect the bombardment on Vesta significantly respect to the fast migration assumed in the reference cases. Concerning the formation of Jupiter, instead, \citet{ward2005} suggested that the combined effects of the resonances due to the planetary embryos accreting to form the Jovian core could affect the asteroid belt and cause its depletion before the giant planet started to capture its gaseous envelope. While the simulations of planetary accretion across the inner and outer Solar System performed by \citet{weidenschilling2008} did not show evidences of such an effect, further and more detailed studies are nevertheless needed to fully assess the dependence of the JEB from the parameters describing the formation and early migration of Jupiter.

\section{Conclusions}

Notwithstanding the uncertainties underlying the physical description of the JEB, which will be addressed in future works, the results presented here clearly highlight that the relative timing of the geophysical evolution of Vesta with the formation of Jupiter, together with the fact that Vesta survived intact and preserved its basaltic crust throughout its whole collisional history, put this asteroid in a unique position to study the first phases of the life of the Solar System. 

While the investigation of the most ancient crater record of the asteroid is severely limited by the saturation of the vestan surface, the quantities studied in this work (the surface erosion, the limited mixing of material from the mantle with that of the crust, the removal of the undifferentiated crust and/or the preservation of the basaltic one) open up new ways to explore the ancient past of Vesta and, with it, the primordial history of the asteroid belt and of the Solar System.

From the results of this work it clearly appears that, if the primordial planetesimals formed in a turbulent Solar Nebula, for Vesta to survive the JEB they needed to have collisionally evolved to a SFD similar to the present one of the asteroid belt (as in the results of \citealt{morbidelli2009}) before Jupiter completed its accretion. In this scenario, the survival of Vesta would constrain the Jovian migration to $0.25$ AU or less. For larger displacements of the giant planet, Vesta would undergo large, basin-forming impacts that would expose the vestan mantle, incompatible with the observational data from the Dawn mission. 

The formation of the primordial planetesimals in a quiescent environment starting from relatively small ($\sim100$ m) building blocks, as in the results of \citet{weidenschilling2011}, appears to be the most favourable scenario for the survival of the basaltic crust of Vesta and, for Jovian displacements up to about $0.5$ AU, is the most compatible with the observational data on the asteroid and the HED meteorites. If, at the time the JEB took place, Vesta possessed a now-lost undifferentiated primordial crust, as recent geophysical and thermal studies suggest, then this scenario would favour displacements of Jupiter comprised between $0.5$ AU and $1$ AU.

Finally, it is worth noting that the excavation of the surface due to impacts and the formation of fractures in the bedrock underlying the craters could have caused both effusive and intrusive phenomena from the internal magma ocean across the whole crust of Vesta. In the latter case, depending on the depth reached by the fractures, intrusions of liquid material from the lower crust (i.e. diogenite) and from the topmost layers of the mantle  (i.e. olivine) could have formed in the solid crust. While speculative, this possibility highlights a link between the mineralogy of HED meteorites and the primordial evolution of the Solar Nebula that deserves to be further investigated in future studies.



\section*{Acknowledgements}

I wish to thank Guy Consolmagno, Hap McSween and Christopher Russell for the discussions on the subjects of the composition and the erosion of the surface of Vesta and Paul Tackley and Gregor Golabek for the discussions on the geophysical evolution of Vesta and its crust. I also want to thank Christopher Russell, Sharon Uy, Romolo Politi, Michelangelo Formisano, Danae Polychroni and the two anonymous reviewers for their comments and assistance in improving the quality of the paper. This research has been supported by the Italian Space Agency (ASI) through the ASI-INAF contract I/010/10/0 and by the International Space Science Institute in Bern through the International Teams 2012 project ``Vesta, the key to the origins of the Solar System'' (http://www.issibern.ch/teams/originsolsys). The computational resources used in this research have been supplied by INAF-IAPS through the project ``HPP - High Performance Planetology''.

\bibliographystyle{elsarticle-harv}



\end{document}